\documentclass[10pt,preprint]{aastex}
\usepackage{graphicx}

\begin{document}

\title{A Galaxy Cluster Near NGC 720}

\author{H. Arp}
\affil{Max-Planck-Institut f\"ur Astrophysik, Karl Schwarzschild-Str.1,
  Postfach 1317, D-85741 Garching, Germany}
 \email{arp@mpa-garching.mpg.de}



\begin{abstract}

The galaxy cluster RXJ 0152.7-1357 is emitting X-rays at the high rate of 148
counts $ks^{-1}$. It would be one of the most luminous X-ray clusters known if it is at 
its redshift distance of z = .8325. It is conspicuously elongated, however, toward the
bright, X-ray active galaxy NGC 720 about 14 arcmin away. At the same
distance on the other side of NGC 720, and almost perfectly aligned, is an X-ray
BSO of 5.8 cts/ks. It is reported here that the redshift of this quasar is z = .8312.

\end{abstract}  

\section{Introduction}

The ESO Messenger (Rosati et al. 2000) carried a color picture of the powerful 
X-ray cluster of galaxies RXJ 0152.7-1357 at z = .83. (Reproduced here in Fig. 1). 
Because the cluster was linearly extended by a ratio of about 3 to 1 Arp looked 
along the direction of the line to see if there was any nearby active galaxy. In fact 
there was a very bright Shapley Ames galaxy, NGC 720 ($B_T$ = 11.15, z = .0057). 
NGC 720 is strongly active with an X-ray filament extending from the nucleus 
and curving southward as it emerges (Buote and Canizares 1996).

The configuration was compelling enough to motivate searching a field of 30'
radius around NGC 720 for additional active objects. The results are shown here
in Fig. 2 (adapted from a picture published in the Catalogue of Discordant Redshift
Associations, Arp 2003). The most conspicuous additional object in the field is 
the rather bright (r, b = 19 mag.) blue stellar object (BSO) which coincides with 
the ROSAT PSPC X-ray source of 5.8 cts/ks. This X-ray BSO is just 14.5' from 
NGC 720. The X-ray galaxy cluster is 14.2' from NGC 720 and the two are 
almost exactly aligned with each other across the galaxy.

\begin{figure}
\includegraphics[width=8.0cm]{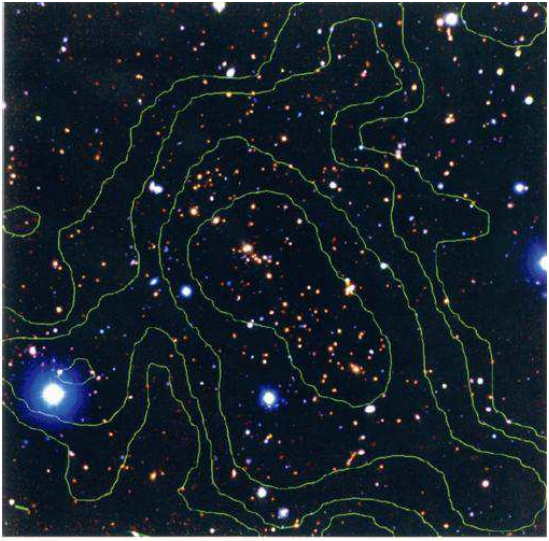}
\caption{An optical image, with X-ray contours superposed, of the cluster 
RXJ0152.7-1357 at z = .833 (Rosati et al. 2000). Frame is 4.7' x 4.7'. At its redshift 
distance it would be one of the most X-ray luminous clusters known. It points, 
however, nearly at the strong X-ray E galaxy NGC 720. (See next Figure.)
\label{fig1}}
\end{figure}

\begin{figure}
\includegraphics[width=9.0cm]{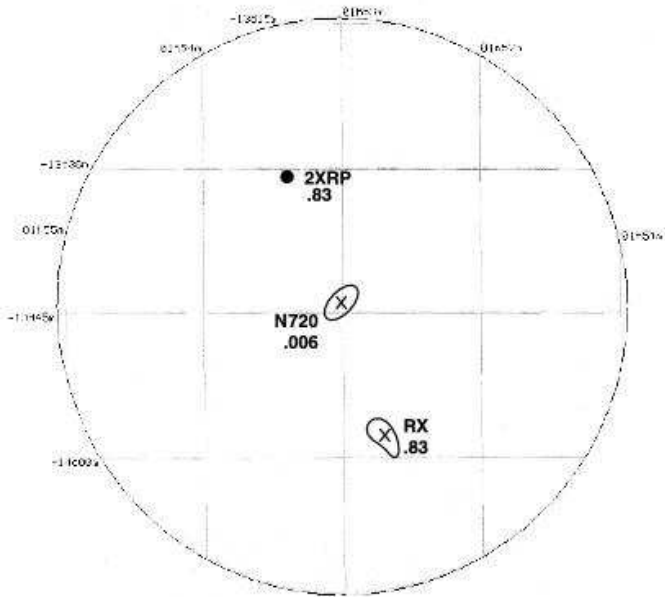}
\caption{Field of 30' radius around the X-ray ejecting NGC 720 (Arp 2003). The
strong X-ray cluster with z = .833 in Fig.1 is shown schematically to the
SW. The spectrum of the X-ray blue stellar object (2XRP) to the NE is shown in 
the next figure to have z = .831.
\label{fig2}}
\end{figure}

The search for telescope time to obtain the spectrum of the candidate BSO was
solved by E. Margaret Burbidge who enabled a 5 minute exposure to be
obtained with the Keck Telescope. The optical spectrum of 2XRP was obtained 
with the Low Resolution Imaging Spectrograph (LRIS) (Oke, Cohen et al. 1995)
attached to the Keck I 10m telescope on Mauna Kea.  We show the
spectrum in Figs. 3 and 4 which identify the major emission
lines. This X-ray/BSO is clearly a QSO with an emission redshift of z = .8312.

The spectrum as shown here in Figs. 3 and 4. is marked by strong
emission MgII and $[CIII]$ in the blue and extraordinarily strong 
$H\beta, H\gamma, H\delta, H\epsilon$ in the red. The latter lines
remind one of strongly star forming galaxies and may contain 
interesting information about the structure of this particular quasar.

\clearpage 

\begin{figure}
\includegraphics[width=18.0cm]{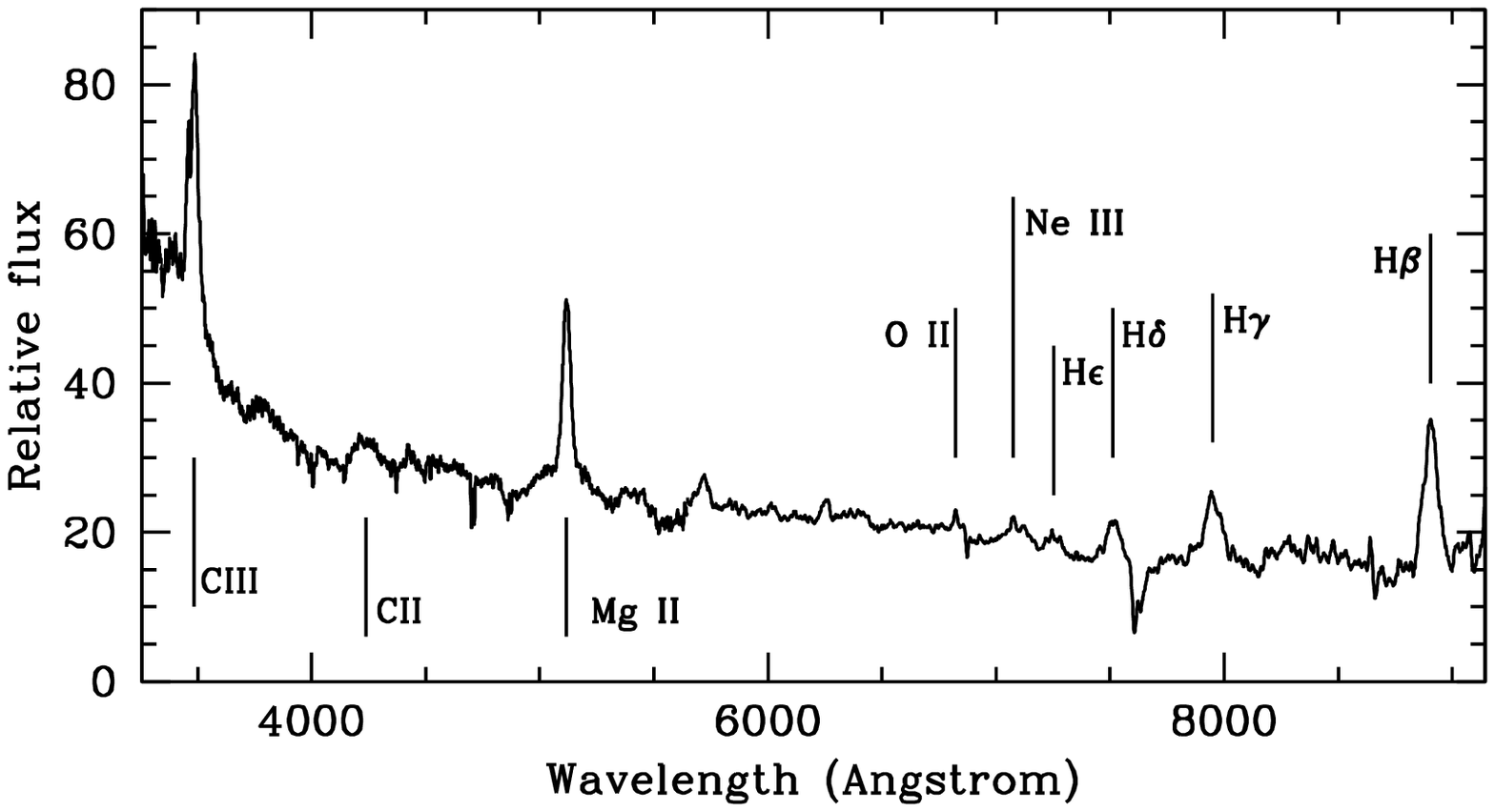}
\caption{A 5 minute spectrum with the Keck telescope on the 2XRP object 14'
north of NGC 720. The CIII] line is measured at 
$\lambda 3485.44$.
\label{fig3}}
\end{figure}

\begin{figure}[h]
\includegraphics[width=8.0cm]{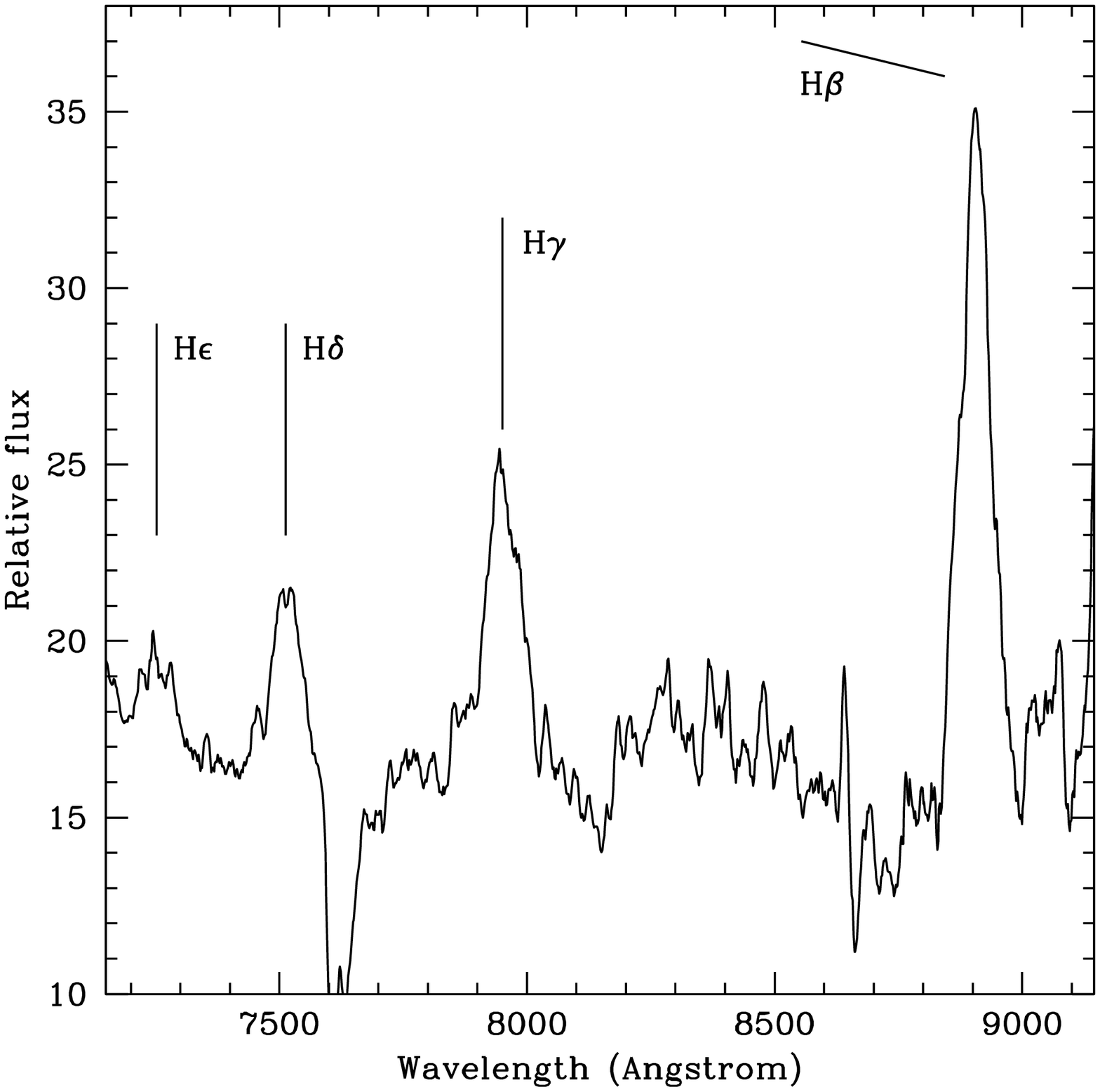}
\caption{An enlargement of the red section of Fig.3 showing the strong Balmer
emission lines in the 2XRP quasar.
\label{fig4}}
\end{figure}

\section{Probability}

The space density of galaxies brighter than $B_T$ =11.15 in the general region 
of NGC 720 is about .009 /sq.deg. To find a galaxy of bright as NGC 720 within 
14' of the X-ray cluster has a probability of about P = .002. But then we have to 
decrease this by at least an order of magnitude in order to find a galaxy as active 
as NGC 720. Moreover there is the additional improbability that the cluster is 
elongated within about 20 deg. back toward NGC 720 and that an X-ray filament 
extends from the nucleus of NGC 720 and curves around toward the south as it 
exits the galaxy (Buote and Canizares 1996).

But now with the strong X-ray/quasar aligned with the strong X-ray cluster 
within a few degrees across NGC 720, and the pair accurately centered,
the improbability of the configuration decreases even more drastically. Finally the
redshifts of the two paired objects are both z = .83 where a random redshift
would be expected somewhere between z = .3 and z = 2.6.

We can estimate the probability of the configuration accidently fulfilling the
prototypical pattern of: active galaxy near bright object, pair of objects aligned
and centered across active galaxy, and similarity of redshifts of paired objects:

$$ 2 x 10^{-4} x 2x 10^{-2} x 5 x 10^{-3} = 10^{-8}$$

Since for each of these properties p = .5 would signify an average random result,
we divide by ${.5^4} = .0625$ giving a final value of:

$$ P = 1.6 x 10^{-7}$$ that the configuration accidently fulfills the empirical 
criteria for ejected objects.

\section{Precedent for Alignment and Ejection}

It is sometimes objected that alignments of quasars and other high red shift 
objects across galaxies are not meaningful because the probability estimates are 
a posteriori. The answer is that the first alignment is a posteriori. Each 
subsequent alignment is confirmation of an a priori prediction. Morover the 
improbabilitiess compound as each new case is discovered. The stated 
characteristics are closeness of bright objects, their alignment, their centering 
and the similarity of the aligned objects. These are all properties expected of 
objects ejected from active galaxies.

The tendency for paired quasars across galaxies to have similar
redshifts is is well documented (e.g. Arp 2003). {\it A previous example of
a pair of almost identical redshift quasars closely paired across
an active galaxy is Arp 220 where the quasar redshifts were z = 1.25
and z = 1.26 (Arp et al. 2001).}
 
The first alignments were established by strong radio quasars (usually
Parks or 3C sources) across peculiar (usually morphologically disturbed
or active nuclei galaxies (Arp 1967; 2003). Because radio sources had been
by then accepted as ejected from galaxy nuclei the aligned high redshift
objects were strongly indicated to be ejecta from the central galaxy.

One of the reasons we know radio sources are ejected is that we observe 
radio emitting gas moving outward in jets that terminate on extended clouds
of radio emission (e.g. Zensus 1997). Since the advent of X-ray astronomy we can 
also observe X-ray winds and jets emerging from active galaxies,
some as a narrow cores to the radio ejections. In the case of pairs and lines
of X-ray sources ejected from active galaxies, however, almost every point X-ray
source can be established as a high redshift quasar. The question is
then posed: Why do many radio sources appear as blank fields with no optical object?

In the past it has been suggested that radio quasars on their way out of
the inner regions of galaxies or through the intergalactic medium are stripped of
their outer layer of lower density, radio emitting gas. (Arp 2001a). In any case the 
more compact X-ray sources are almost always optically identified,
form an even more consistent pattern of pairing 
across active galaxies and empirically support the picture of ejection of high red 
objects. (See samples of the many cases as reviewed in  Arp (1987; 1998) and 
Catalog of Discordant Redshift Associations (Arp 2003.)

\section{Evidence for Association of Galaxy Clusters with Low Redshift Galaxies}

It is almost universally believed that galaxy clusters are distant aggregates of
normal, luminous galaxies and that they could not possibly be associated with 
nearby, low redshift progenitors. There is, however, considerable empirical
evidence that the latter is the case. For example the very bright, nearby, radio 
and X-ray 
ejecting galaxy, Cen A (NGC 5128) was shown to have a cone of higher redshift 
galaxy clusters extending 15 - 20 deg. northward along the direction of its radio 
ejection (Arp 1998, p147). Later a pair of X-ray clusters was shown across 
the X-ray galaxy ESO 185-54. The SE of this pair, A3667 was a very strong in 
X-rays at 2440 cts/ks. It was elongated back toward the central galaxy and 
the low and high redshift galaxies were intermixed along the connection. 
{\it Moreover high resolution Chandra observations captured a bow shock at the 
head of A3667 which showed the it was moving with 1400 km/sec directly away 
from the low redshift, central, X-ray galaxy down the extended line of
mixed redshift galaxies (Arp 2003)}

Further evidence for a number of cases of Abell clusters physically associated
with bright, low redshift, nearby galaxies was presented by Arp and Russell 
(2001). One point made in that publication was that at the nearer 
distances for these clusters, the galaxies would be moderately to somewhat 
underluminous but at their conventional redshift distance they would be 
unprecedentedly over luminous .

\section{Are proto Clusters of Galaxies Ejected like quasars?}

If we accept that both quasars and galaxy clusters have their origin in
ejections from low redshift galaxies we must face the question of what is their 
relation to each other. We start with the question of what gives the quasars their
intrinsic redshifts. Here we invoke the variable mass hypothesis of Narlikar (1977)
and Narlikar and Arp (1993). In that solution of the general relativistic field
equations the particle masses of new matter start out at or near zero mass
and grow with time. Because the electrons making orbital transitions in radiating
atoms are initially small the emitted photons are initially redshifted and decrease
their intrinsic redshift with time. The quasars are then viewed as being composed
of young matter which evolves toward normal matter and normal galaxies with 
time.

In the initially ejected proto quasars the particles grow in mass and slow down in 
order to conserve momentum so the particles cool and increasingly gravitate 
toward a young galaxy (no dark matter needed). The initial plasmoid, however, 
has low mass ions which have large cross sections and are more strongly 
frozen in by magnetic fields (Arp 1963). In order to have the time to evolve  
intrinsic redshifts into the range of older galaxies like our own they must be 
slowed down or stopped by the passage through the internal regions of the 
parent galaxies or meeting clouds in the medium exterior to the galaxy. They 
should be fragile and observations suggest that some quasars are split 
into two's or three's (Arp and Russell 2001, p548; Arp 1999). In practice it 
is suggested that sometimes they can run into a medium of cloudlets and be 
divided into many small proto galaxies. i.e. a cluster of proto galaxies on its way 
to evolving into a cluster of galaxies. 

It is worth noting that active galaxies, quasars and clusters of galaxies are the 
three principle kinds of extra galactic X-ray sources that exist. In the above 
picture each are subunits of the former. The processes in galaxy nuclei which 
give rise to the quasars furnish the energy to fission or explode some quasars 
into smaller pieces which evolve into galaxy clusters particularly in interaction
with a  galaxy/extra galactic environment. An example of a group of X-ray 
quasars which should evolve into a galaxy cluster are those roughly aligned
across 3C345 (Arp 1997).

Aside from discussions about overluminous galaxies at conventional
distances versus underluminous galaxies in nearby clusters, there are
the large cluster distances calculated on the assumptions of the
Sunyaev-Zeldovich effect. But as discussed in the Moriond presentation Arp 
(2001b) the non-equilibrium physical state of low particle mass plasmas may
give much different predictions in the S-Z analysis than in the present models.

\section{Summary}

The preceding qualitative discussion of theory is intended to suggest a 
physically plausible model which could explain the empirical evidence for the 
relation between quasars and galaxy clusters of various redshifts. The 
overridingly important point, however is that the NGC 720 case reported here 
confirms at a very high level of significance the previous evidence for ejection of 
much higher redshift objects from active galaxy nuclei. It would seem that the 
recognition of this evidence should necessitate a reconsideration of the 
fundamental assumptions about extragalactic redshifts.

A more conventional interpretation of this unusual grouping of X-ray sources is
in press in the Publ. Astron. Soc. Pacific (E.M. Burbidge, H.C. Arp 2005).

\end{document}